\documentclass[journal]{IEEEtran}

\usepackage{cite}

\ifCLASSINFOpdf
   \usepackage[pdftex]{graphicx}
   \graphicspath{{./figures}}
  \DeclareGraphicsExtensions{.pdf,.jpeg,.png,.eps}
\else
\fi

\usepackage{amsthm}
\usepackage{amsmath}
\usepackage{cases}
\usepackage{siunitx}
\usepackage{cleveref}

\crefname{equation}{equation}{}
\Crefname{equation}{Equation}{Equations}
\crefrangelabelformat{equation}{(#3#1#4--#5#2#6)}

\crefmultiformat{equation}{(#2#1#3}{, #2#1#3)}{#2#1#3}{#2#1#3}
\Crefmultiformat{equation}{Equations (#2#1#3}{, #2#1#3)}{#2#1#3}{#2#1#3}

\usepackage{color}

\usepackage{algorithmic}

\usepackage{array}
\usepackage{multirow}

\ifCLASSOPTIONcompsoc
  \usepackage[caption=false,font=normalsize,labelfont=sf,textfont=sf]{subfig}
\else
  \usepackage[caption=false,font=footnotesize]{subfig}
\fi

\usepackage{url}

\usepackage{subfig}

\usepackage{arydshln}

\begin{document}

\newcounter{MYtempeqncnt}
\newtheorem{theorem}{Theorem}

\title{Feedback Linearization for Quadrotors UAV}

\author{Mauricio~Alejandro~Lotufo, Luigi~Colangelo, and~Carlo~Novara,~\IEEEmembership{Senior Member,~IEEE}
\thanks{M. A. Lotufo is with the Politecnico di Torino, Torino, 10129 Italy (e-mail: mauricio.lotufo@polito.it).}
\thanks{L. Colangelo is with the Department of Electronics and Telecommunications Engineering, Politecnico di Torino, Torino, 10129 Italy (e-mail: luigi.colangelo@polito.it).}
\thanks{C. Novara is with the Department of Electronics and Telecommunications Engineering, Politecnico di Torino, Torino, 10129 Italy (e-mail: carlo.novara@polito.it).}
}%



\maketitle
\IEEEpeerreviewmaketitle

\begin{abstract}
In the paper ``\textit{Control Design for UAV Quadrotors via Embedded Model Control}"~\cite{lotufoIEEE_TBC}, the authors designed a complete control unit for a UAV Quadrotor, based on the Embedded Model Control (EMC) methodology, in combination with the Feedback Linearization (FL); when applied to non-linear systems. Specifically,~\cite{lotufoIEEE_TBC} proposes to use the FL as a novel way to design the internal model for the EMC state and disturbance predictor. To support the treatise in~\cite{lotufoIEEE_TBC}, in this report the feedback-linearized model of the UAV quadrotor leveraged in~\cite{lotufoIEEE_TBC} is step-by-step derived.
\end{abstract}

\IEEEpeerreviewmaketitle

	\section{Introduction}
	\label{sec:intro}
\IEEEPARstart{F}{eedback} Linearization (FL) technique allows transforming a command-affine non-linear model of the UAV quadrotor into an equivalent (fully or partly) linear one. Specifically, FL: (i) pursues the collection of all the model non-linearities in specific points, e.g. at the command level, and (ii) achieves an input-output linearization by means of a non-linear feedback, performing a perfect cancellation of non-linearities~\cite{khalil1996noninear}.

Nevertheless, model uncertainties make non-linear terms uncertain, and their use into the FL feedback may imply performance degradation and/or instability. Hence, this study proposes to use FL in combination with the Embedded Model Control (EMC) framework. In short, the designed FL-EMC approach let us to treat non-linearities as known and unknown disturbances to be estimated and then rejected, thus enhancing control robustness and performance. To this purpose, the study in~\cite{lotufoIEEE_TBC} focused on the so-called normal form representation of the non-linear model, where the non-linearities are collected at the command level, which perfectly fits the EMC internal model design rationale~\cite{lotufoIEEE_TBC}.

	\section{Feedback Linearization}
	\label{sec:fla}
The feedback linearization (FL) technique is an effective resource to linearize a non-linear model, by introducing a proper state transformation and a non-linear feedback~\cite{slotine1991applied}. Then, starting from the new linear model (i.e. the feedback-linearized one), a linear controller can be designed. In practice, within the FL, the model input-output linearization is obtained by differentiating each model output many times, until a control input component appears in the resulting equation.

Generally speaking, the feedback-linearized model is obtained by means of a system state transformation (diffeomorphism) and a non-linear feedback~\cite{slotine1991applied}. The state variables of the transformed model are the Lie derivatives of the system output $\mathbf{y}$. This implies that the choice of the output vector is extremely important to accomplish the input-output linearization.

Let us consider a command-affine square non-linear system, with state vector $\mathbf{x}\,{\in}\,R^{n}$, input $\mathbf{u}\,{\in}\,R^{m}$, and output $\mathbf{y}\,{\in}\,R^{m}$:
	\begin{equation}
	\label{eq:model_nl}
	\begin{split}
	\dot{\mathbf{x}}(t) &= \mathbf{f}(\mathbf{x}(t)) + G(\mathbf{x}(t))\mathbf{u}(t) , \: \mathbf{x}(0)=\mathbf{x}_0, \\
	\mathbf{y}(t) &= \mathbf{h}(\mathbf{x}(t)),	
	\end{split}
	\end{equation}
where $\mathbf{f}$ and $\mathbf{h}$ represent smooth vector fields~\cite{khalil1996noninear}, while $G\,{\in}\,R^{{n{\times}m}}$ is a matrix with smooth vector fields as columns. Denoting with $r_i$ the relative degree of the $i{-}th$ output, we aimed to obtain an equivalent non-linear model, where all non-linearities have been collected at the command level, i.e.:
	\begin{equation}
	\label{eq:model_fl}
	\begin{bmatrix}
	y_1^{(r_1)} \\ y_2^{(r_2)} \\ \dots \\ y_m^{(r_m)} \\
	\end{bmatrix}(t) = E(\mathbf{x}(t))\mathbf{u}(t) + \mathbf{b}(\mathbf{x}(t)),
	\end{equation}
where $y^{(n)}(t)$ denotes the time derivative of order $n$. Specifically,~\eqref{eq:model_fl} represents a cascade of integrators in parallel, where the output and its derivatives are the new state variables $\mathbf{z}$ defined as:
	\begin{equation}
	\label{eq:statetrans}
	\begin{split}
	\mathbf{z}\,{=}\,T(\mathbf{x})\,{=}\,[	& y_1 \:\: y_1^{(1)} \:\: \dots \:\: y_1^{(r_1-1)} \:\: \dots \\ 
	 										& y_2 \:\: y_2^{(1)} \:\: \dots \:\: y_2^{(r_2-1)} \:\: \dots \\ 
	 										& \dots \\
	 										& y_m \:\: y_m^{(1)} \:\: \dots \:\: y_m^{(r_m-1)} ]^T,
	\end{split}
	\end{equation}
where $T(\mathbf{x})$ represents a diffeomorfism.
Whenever this state transformation is applicable and the decoupling matrix $E(\cdot)$ is invertible, it is possible to linearize the equivalent model~\eqref{eq:model_fl} via the non-linear feedback:
	\begin{equation}
	\label{eq:law_fl}
	\begin{split}
	\mathbf{u}(t) = E(\mathbf{x}(t))^{-1} \left( \mathbf{v}(t) - \mathbf{b}(\mathbf{x}(t)) \right).
	\end{split}
	\end{equation}
where $\mathbf{v}$ is a new equivalent command. Hence, by applying the feedback~\eqref{eq:law_fl} to the model~\eqref{eq:model_fl}, a parallel of $m$ decoupled input-output channels, represented by cascaded integrators, is obtained, viz.:
	\begin{equation}
	\label{eq:lin_mdl}
	\begin{split}
	\begin{bmatrix}
	y_1^{(r_1)} \\ y_2^{(r_2)} \\ \dots \\ y_m^{(r_m)} \\
	\end{bmatrix}(t) &= \mathbf{v}(t)
	\end{split},
	\end{equation}
where the new command $\mathbf{v}$ is used for the design of a linear controller.

A ``full" input-output linearization is achieved if the sum of the output relative degrees is equal to the order of the model~\eqref{eq:model_nl}~\cite{slotine1991applied}. When this condition is not verified, some dynamics are hidden by the state transformation. This so-called ``internal" dynamics could be unstable~\cite{slotine1991applied}. In this case the feedback linearization process fails.

Finally, when $E(\cdot)$ is not invertible in $R^{n}$, it is possible to adopt a new command vector by considering the derivative of some of the command components. This solution, called dynamic extension~\cite{slotine1991applied}, enables the applicability of the feedback linearization by making the $E(\cdot)$ invertible, yet at a cost: the introduction of a dynamics in the command. 
Furthermore, let us remark that the non-singularity condition of $E(\cdot)$ is not necessary verified in the complete state space. As a matter of fact, in some applications, these singularities can be avoided by applying proper constrains to the state trajectories.

	\section{The Quadrotor UAV Case-study}
	\label{sec:quad_fla}
As above depicted, the input-output linearization is achieved by differentiating each output many times until a control input appears. Hence, the successful application of the feedback linearization is strongly dependent on the output vector.

As a matter of fact, the adopted Quadrotor UAV dynamics, which encompasses four commands and three outputs~\cite{lotufoIEEE_TBC}, is characterized by a non-square decoupling matrix $E(\cdot)\,{\in}\,\mathcal{R}^{3,4}$~\cite{mistler2001_FL}. Therefore, in~\cite{lotufoIEEE_TBC}, the quadrotor heading angle $\psi$ was selected as an additional output to be controlled, so to make the FL applicable to the Quadrotor UAV model~\cite{lotufoIEEE_TBC}. 

\begin{figure}[!t]
	\centering
	\includegraphics[width=2.5in]{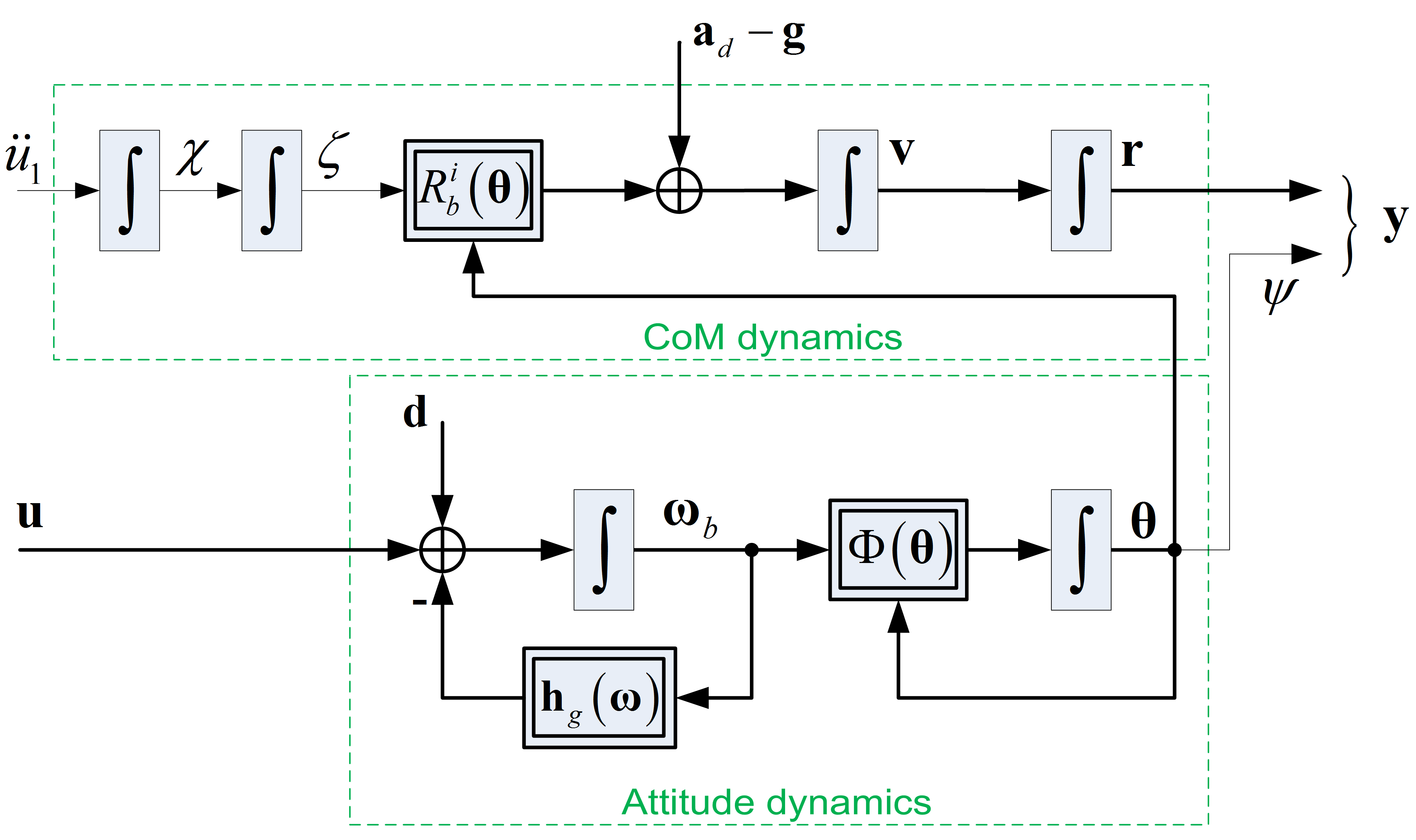}
	\caption{Quadrotor UAV extended Model.}
	\label{fig:extended}
	\end{figure}

However, in the quadrotor UAV case-study treated in~\cite{lotufoIEEE_TBC}, the problem is not solvable by means of a static feedback~\cite{mistler2001_FL}. Conversely, a full linearization may be only achieved by introducing additional states to the standard quadrotor model, thus obtaining the so-called extended model~\cite{mistler2001_FL}. In fact, in order to obtain a non-singular decoupling matrix $E(\cdot)$, the second derivative of the first command component, $u_1$ in \figurename~\ref{fig:extended}, with respect to time was also needed.

As a result, the introduction of two more states, i.e. $\zeta\,{=}\,u_1$ and its derivative $\chi\,{=}\,\dot{u}_1$, lead to define the Quadrotor UAV extended model, whose state vector is $\mathbf{x}\,{=}\,\left[ \mathbf{r}^T \:\: \mathbf{v}^T \:\: \boldsymbol{\theta}^T \:\: \boldsymbol{\omega}_b^T \:\: \zeta \:\: \chi \right]^T$, where $\mathbf{r}$ and $\mathbf{v}$ are respectively the inertial position and velocity of the CoM, $\boldsymbol{\theta}=[\phi \:\: \theta \:\: \psi]^T$ are the attitude angles, and $\boldsymbol{\omega}_b$ is the body angular rate vector. 
Hence, the model, sketched in \figurename~\ref{fig:extended}, holds~\cite{lotufoIEEE_TBC}:
	\begin{equation}
	\label{eq:extended_quad}
	\begin{split}
	\boldsymbol{\dot{\theta}}(t) &= A(\boldsymbol{\theta}(t)) \boldsymbol{\omega}_b(t), \quad \boldsymbol{\theta}(0)=	\boldsymbol{\theta}_0, \\
	A(\boldsymbol{\theta}) &=
	\frac{1}{c_{\theta}} 
	\begin{bmatrix}  
	c_{\psi} & -s_{\psi} & 0 \\
	c_{\theta}s_{\psi} & c_{\theta}c_{\psi} & 0 \\
	-s_{\theta}c_{\psi} & s_{\theta}s_{\psi} & c_{\theta}   
	\end{bmatrix}, \\	
	\dot{\boldsymbol{\omega}}_b(t) &= \mathbf{u}(t) - J^{-1}(\boldsymbol{\omega}_b(t) \times J\boldsymbol{\omega}_b(t)) + \mathbf{d}(t), \\
	\boldsymbol{\omega}_b(0) &= \boldsymbol{\omega}_{b0}, \\
	\dot{\mathbf{r}}(t) &= \mathbf{v}(t), \quad \mathbf{r}(0) = \mathbf{r}_0, \\
	\dot{\mathbf{v}}(t) &= R_{b}^i(\boldsymbol{\theta})\begin{bmatrix} 0 & 0 & \zeta(t) \end{bmatrix}^T
	- \mathbf{g} + \mathbf{a}_d(t), \\
	\mathbf{v}(0) &= \mathbf{v}_0, \\
	\dot{\zeta}(t) &= \chi(t), \\
	\dot{\chi}(t) &= \ddot{u}_1(t), \\
	\mathbf{y}(t) &= \begin{bmatrix} \mathbf{r} \\ \psi \end{bmatrix}(t), \quad
	\overline{\mathbf{u}} = \begin{bmatrix} \ddot{u}_1 \\ \mathbf{u} \end{bmatrix}(t).
	\end{split}
	\end{equation}
In~\eqref{eq:extended_quad}, $\mathbf{u}$ is the command torque along the three body axes, while $J$ is the quadrotor inertia matrix. In addition, $R^i_{b}(\boldsymbol{\theta})$ describes the body-to-inertial attitude, $\mathbf{g}\,{=}\,\left[ 0 \:\: 0 \:\: 9.81 \right]^T$ is the gravity vector, while $\mathbf{d}$ and $\mathbf{a}_d$ represent all the external disturbances (e.g. wind, rotor aerodynamics, mechanical vibration, actuator noise) affecting the model dynamics. Finally, $\overline{\mathbf{u}}$ and $\mathbf{y}$ are the new command vector and the extended model output, respectively.

The extended model~\eqref{eq:extended_quad} is the foremost building block of the feedback linearization process, performed to obtain the feedback-linearized model leveraged in~\cite{lotufoIEEE_TBC} to design the internal model for the EMC state and disturbance predictor. To this aim, starting from the overall model in~\eqref{eq:extended_quad}, we will derive the feedback-linearized heading and CoM dynamics, through the procedure outlined in Sec. II.

\subsection{The Quadrotor UAV Dynamics}
	\label{subsec:quad_fla_head}
Considering the quadrotor attitude kinematics in~\eqref{eq:extended_quad}, let us start from the heading dynamics. The first order time-derivative of the heading angle holds:
	\begin{equation}
	\label{eq:psi_1}
	\begin{split}
	\psi^{(1)}(t) &= \eta(t) = \mathbf{b}_{\psi}(\boldsymbol{\theta}(t))\boldsymbol{\omega}_b(t), \\
	\mathbf{b}_{\psi}(\boldsymbol{\theta}(t)) &= [ -t_\theta c_\psi \quad t_\theta s_\psi \quad 1].
	\end{split}
	\end{equation} 
A further derivative is needed in order to relate the output with the command, viz.:
	\begin{equation}
	\label{eq:fl_heading}
	\begin{split}
	\psi^{(2)}(t) &= \mathbf{b}_{\psi}(\boldsymbol{\theta}(t))(\mathbf{u}(t) - \mathbf{h}_g(t) + \mathbf{d}(t)) + 	\dot{\mathbf{b}}_{\psi}(\boldsymbol{\theta}(t))\boldsymbol{\omega}_b(t)= \\
	&= \mathbf{b}_{\psi}(\boldsymbol{\theta}(t))\mathbf{u}(t) + h_\psi(\mathbf{x}(t)) + d_\psi(t),
	\end{split}
	\end{equation}
where we defined $h_\psi$ and $d_\psi$ as:
	\begin{equation}
	\label{eq:fl_heading2}
	\begin{split}
	h_\psi(\mathbf{x}(t)) &= \dot{\mathbf{b}}_{\psi}(\boldsymbol{\theta}(t))\boldsymbol{\omega}_b(t) - \mathbf{b}_{\psi}	(\boldsymbol{\theta}(t))\mathbf{h}_g(t), \\
	d_\psi(t) &= \mathbf{b}_{\psi}(\boldsymbol{\theta}(t))\mathbf{d}(t).
	\end{split}
	\end{equation}

Concerning the CoM dynamics, to relate the output with the command (cf. Sec.~\ref{sec:fla}), two derivatives will be needed; starting from the CoM acceleration.

Specifically, the CoM position third derivative, i.e. the jerk $\mathbf{s}$, holds:
	\begin{multline}
	\label{eq:jerk}
	\mathbf{r}^{(3)}(t) = \mathbf{s}(t) = \\
	= R_b^i(\boldsymbol{\theta}(t))\left( \begin{bmatrix}
	0 \\ 0 \\ \chi(t)
	\end{bmatrix} +
	S(\boldsymbol{\omega}_b(t))\begin{bmatrix} 0 \\ 0 \\ \zeta(t) \end{bmatrix} \right) + \dot{\mathbf{a}}_d(t),
	\end{multline}
being $S(\boldsymbol{\omega}_b)$:
	\begin{equation}
	\label{eq:jerk2}
	\begin{split}
	S(\boldsymbol{\omega}_b) = \begin{bmatrix}
	0 & -\omega_z & \omega_y \\ \omega_z & 0 & -\omega_x \\ -\omega_y & \omega_x & 0
	\end{bmatrix}
	\end{split}.
	\end{equation}
In~\eqref{eq:jerk}, the first term on the right-hand side represents the contribution of the vertical jerk command, whereas the second term is the jerk component due to the quadrotor angular rates.

Moreover, a further derivative of~\eqref{eq:jerk} is necessary in order to have a full-rank decoupling matrix $E(\cdot)$. Hence, it holds:
	\begin{multline}
	\label{eq:r4_1}
	\mathbf{r}^{(4)}(t) = R_b^i(\boldsymbol{\theta}(t))\begin{bmatrix}
	\dot{\omega}_y \zeta \\ -\dot{\omega}_x \zeta \\ \ddot{u}_1 \end{bmatrix}(t) + \\
	+ 2\chi(t)R_b^i(\boldsymbol{\theta}(t))\begin{bmatrix}
	\omega_y \\ -\omega_x \\ 0 \end{bmatrix}(t) + \\
	+ \zeta(t)R_b^i(\boldsymbol{\theta}(t))\begin{bmatrix}
	\omega_x\omega_z \\ \omega_y\omega_z \\ -(\omega_x^2 + \omega_y^2)
	\end{bmatrix}(t) + \ddot{\mathbf{a}}_d(t).
	\end{multline}
As a result, by introducing the quadrotor attitude dynamics from~\eqref{eq:extended_quad} in~\eqref{eq:r4_1}, we obtain:
	\begin{equation}
	\label{eq:fl_com}
	\begin{split}
	\mathbf{r}^{(4)}(t) &= R_b^i(\boldsymbol{\theta}(t))\begin{bmatrix}
	\zeta u_3 \\ - \zeta u_2 \\ \ddot{u}_1
	\end{bmatrix}(t) + \mathbf{h}_r(\mathbf{x}(t)) + \mathbf{d}_r(\mathbf{x}(t),t),
	\end{split}
	\end{equation}
where $\mathbf{h}_r$ and $\mathbf{d}_r$ are the known and the unknown disturbance terms, respectively. Specifically, $\mathbf{h}_r(\mathbf{x}(t))$ was defined as:
	\begin{multline}
	\label{eq:fl_com2}
	\mathbf{h}_r(t) = \zeta(t)R_b^i(\boldsymbol{\theta}(t))\begin{bmatrix}
	\omega_x\omega_z -h_{gy} \\ \omega_y\omega_z + h_{gx} \\ - (\omega_x^2 + \omega_y^2)
	\end{bmatrix}(t) + \\
	+ 2\chi(t)R_b^i(\boldsymbol{\theta}(t))\begin{bmatrix}
	\omega_y \\ -\omega_x \\ 0 \end{bmatrix}(t),
	\end{multline}
whereas $\mathbf{d}_r(\mathbf{x}(t))$ holds:
	\begin{equation}
	\label{eq:fl_com_d}
	\begin{split}
	\mathbf{d}_r(\mathbf{x}(t)) = R_b^i(\boldsymbol{\theta}(t))\begin{bmatrix}
	\zeta d_{ry} \\ -\zeta d_{rx} \\ 0
	\end{bmatrix}(t) + \ddot{\mathbf{a}}_d(t).
	\end{split}
	\end{equation}
From~\eqref{eq:fl_com_d}, it is interesting to notice how only tilt disturbances ($d_{rx}$, $d_{ry}$) act on the CoM dynamics, and how they are amplified by the body vertical acceleration $\zeta\,{=}\,u_1$, which is close to the gravity value for non aggressive manoeuvres~\cite{lotufo2016}.

	\subsection{The Quadrotor UAV Case-study: Complete Model}
	\label{subsec:quad_fla_head}
As a further step, putting together the heading dynamics~\eqref{eq:fl_heading} and the CoM dynamics~\eqref{eq:fl_com}, the whole input-output relation is found:
	\begin{multline}
	\label{eq:fl_final}
	\begin{bmatrix} \mathbf{r}^{(4)} \\ \psi^{(2)} \end{bmatrix}(t) =
	E(\boldsymbol{\theta}(t),\zeta(t))\overline{\mathbf{u}}(t) + \begin{bmatrix} \mathbf{h}_r \\ h_\psi \end{bmatrix}(t) + 
	\begin{bmatrix} \mathbf{d}_r \\ d_\psi \end{bmatrix}(t),
	\end{multline}
where
	\begin{equation}
	\label{eq:fl_final2}
	\begin{split}
	E(\boldsymbol{\theta}(t),\zeta(t)) &= \left[ \begin{array}{c;{2pt/2pt}c}
	B_r(\boldsymbol{\theta}(t),\zeta(t)) & \begin{array}{c}
	0 \\ 0 \\ 0
	\end{array} \\ \hdashline[2pt/2pt]
	\begin{array}{ccc}
	0 & -t_\theta c_\psi & \quad t_\theta s_\psi
	\end{array} & 1
	\end{array} \right] \\	
	B_r(\boldsymbol{\theta}(t),\zeta(t)) &= R_b^i(\boldsymbol{\theta}(t))\begin{bmatrix}
	0 & 0 & \zeta(t) \\ 0 & -\zeta(t) & 0 \\ 1 & 0 & 0 \end{bmatrix}.
	\end{split}
	\end{equation}
As shown in~\eqref{eq:statetrans}, the state vector of the new equivalent model~\eqref{eq:fl_final} is defined by $\mathbf{z}\,{=}\,T(\mathbf{x})\,{=}\,[\mathbf{r} \:\: \mathbf{v} \:\: \mathbf{a} \:\: \mathbf{s} \:\: \psi \:\: \eta]^T$. 
More interestingly, the total relative degree of the model in~\eqref{eq:fl_final} is equal to the order of the extended model in~\eqref{eq:extended_quad}, namely $r_1\,{+}\,r_2\,{+}\,r_3\,{+}\,r_4\,{=}\,n\,{=}\,14$: this implies that no internal dynamics exists, and a full input-output linearization has been achieved.

Furthermore, the decoupling matrix $E(\mathbf{x}(t))$ is non-singular in $D\,{=}\,\lbrace \mathbf{x}(t) \subset R^n : \zeta(t) \neq 0, |\phi(t)|<\pi/2, |\theta(t)|<\pi/2 \rbrace $. This implies that aggressive manoeuvres may be performed, although with tilt angles lower than $\pi/2$ (acrobatic manoeuvres, such as 360-loops, were not in the scope of~\cite{lotufoIEEE_TBC}). On the other side, since the actuators effect is lower-bounded by a minimum saturation thrust, the total vertical acceleration is always positive. For these practical reasons, the state trajectories was constrained to the domain $D$, where the invertibility of the decoupling matrix is guaranteed. 

\begin{figure}[!t]
	\centering
	\includegraphics[width=2.5in]{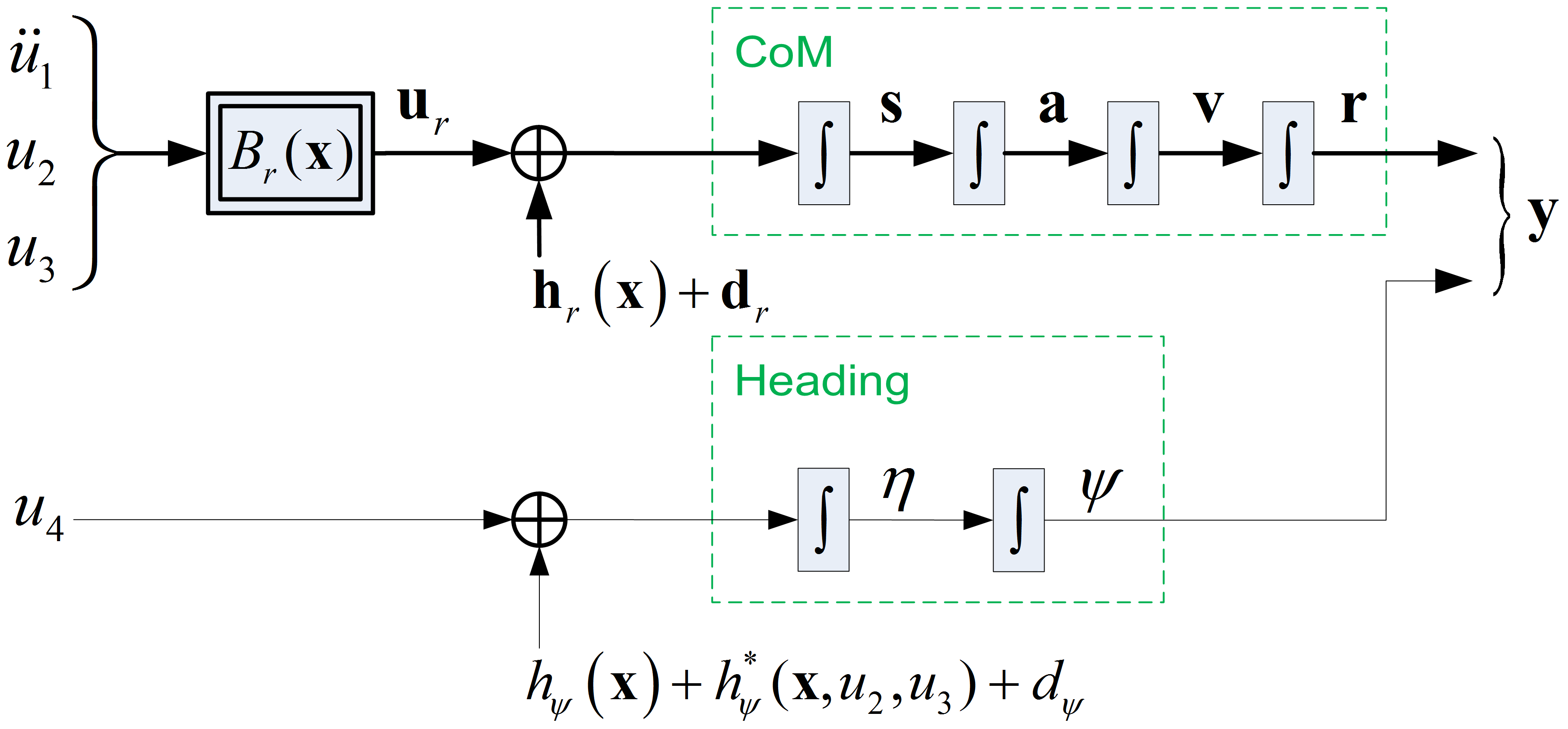}
	\caption{Global scheme of the quadrotor UAV input-output linearized model, as result of the FL technique (Courtesy: \cite{lotufoIEEE_TBC}).}
	\label{fig:model}
	\end{figure}

To conclude, \figurename~\ref{fig:model} sketches the final model~\eqref{eq:fl_final} (cf. also~\cite{lotufoIEEE_TBC}), where the non-linear couplings with the commands $u_2$ and $u_3$ in $\mathbf{b}_\psi$ were collected in $h_\psi^*(\mathbf{x}(t),u_2(t),u_3(t))\,{=}\,-t_\theta c_\psi u_2(t)\,{+}\,t_\theta s_\psi u_3(t)$. On the other hand, $\mathbf{h}_r$ and $h_\psi$ collect all the non-linearities, collocated at command level. Finally, consistently with the EMC design framework, the terms $\mathbf{d}_r$ and $d_\psi$ represent the non-explicitly modelled effects and the external disturbances.

As per \figurename~\ref{fig:model}, for the purpose of the linear control design, the model~\eqref{eq:fl_final} can be rewritten as:
	\begin{subequations}
	\label{eq:fl}
	\begin{align}
	\dot{\mathbf{r}}(t) &= \mathbf{v}(t), \quad \mathbf{r}(0) = \mathbf{r}_0, \label{eq:flA1} \\
	\dot{\mathbf{v}}(t) &= \mathbf{a}(t), \quad \mathbf{v}(0) = \mathbf{v}_0, \label{eq:flA2} \\
	\dot{\mathbf{a}}(t) &= \mathbf{s}(t), \quad \mathbf{a}(0) = \mathbf{a}_0, \label{eq:flA3} \\
	\dot{\mathbf{s}}(t) &= \mathbf{u}_r(t) + \mathbf{h}_r(\mathbf{x}(t)) + \mathbf{d}_r(t), \quad \mathbf{s}(0) = \mathbf{s}_0, \label{eq:flA4} \\
	\dot{\psi}(t) &= \eta(t), \quad \psi(0) = \psi_0, \label{eq:flB1} \\
	\dot{\eta}(t) &= u_4(t) + h_\psi(\mathbf{x}(t)) + h_\psi^*(\cdot) + d_\psi(t), \quad \eta(0) = \eta_0, \label{eq:flB2}
	\end{align}
	\end{subequations}
where $\mathbf{u}_r$ is a transformed command, defined as:
	\begin{equation}
	\label{eq:new_ur}
	\begin{split}
	\mathbf{u}_r(t) &= \mathbf{B}_r(\mathbf{x}(t)) \begin{bmatrix} \ddot{u}_1 & u_2 & u_3 \end{bmatrix}^T(t).
	\end{split}
	\end{equation}
As a result,~\cref{eq:flA1,eq:flA2,eq:flA3,eq:flA4,eq:flB1,eq:flB2} represent the UAV quadrotor model where all the non-linearities have been collocated at the command level and therefore can be cancelled by a non-linear feedback in the form expressed in~\eqref{eq:law_fl}. Nevertheless, this approach relies on the perfect knowledge of the model non-linearities ($\mathbf{h}_r$ , $h_{psi}$) which may considerably limit the controller performance as well as its practical applicability. 

To this aim, the novel approach proposed in~\cite{lotufoIEEE_TBC} (namely, the FL-EMC design) considers the model~\eqref{eq:fl} from a different point of view. More precisely, the non-linear components are treated as generic unknown disturbances which are real-time estimated by a proper extended state observer. Thus, implementing a direct disturbance rejection, jointly with a linear control law, it is possible to completely neglect model non-linearities and, at same time, to enhance the controller robustness against model uncertainties.




%

\ifCLASSOPTIONcaptionsoff
  \newpage
\fi

\bibliographystyle{IEEEtran}

\bibliography{./bib/library,./bib/quadrotor_library,./bib/library_paper}

\end{document}